\documentclass[aps,prl,twocolumn,superscriptaddress]{revtex4}
\usepackage{graphicx}
\begin{document}
\title{Effect of Cu doping on superconductivity in TaSe$_3$: Relationship between superconductivity and induced charge density wave}
\author{Atsushi Nomura}
\affiliation{Department of Applied Physics, Hokkaido University, Sapporo 060-8628, Japan.}

\author{Kazuhiko Yamaya}
\affiliation{Center of Education and Research for Topological Science and Technology, Hokkaido University, Sapporo 060-8628, Japan.}

\author{Shigeru Takayanagi}
\affiliation{Center of Education and Research for Topological Science and Technology, Hokkaido University, Sapporo 060-8628, Japan.}

\author{Koichi Ichimura}
\affiliation{Department of Applied Physics, Hokkaido University, Sapporo 060-8628, Japan.}
\affiliation{Center of Education and Research for Topological Science and Technology, Hokkaido University, Sapporo 060-8628, Japan.}

\author{Satoshi Tanda}
\affiliation{Department of Applied Physics, Hokkaido University, Sapporo 060-8628, Japan.}
\affiliation{Center of Education and Research for Topological Science and Technology, Hokkaido University, Sapporo 060-8628, Japan.}


\begin{abstract}
By measuring the temperature dependence of the resistance, we investigated the effect of Cu doping on superconductivity (SC) in Cu-doped TaSe$_3$ in which the charge density wave (CDW) transition is induced by Cu doping. We observed an emergence of a region where the SC transition temperature ($T_\mathrm{C}$) decreased in samples with higher Cu concentrations and found that the region tended to expand with increasing Cu concentration. In addition, the temperature dependence of the upper critical field ($H_\mathrm{C2}$) of Cu-doped TaSe$_3$ was found to differ from that of pure TaSe$_3$. Based on these experimental results and the fact that the SC of TaSe$_3$ is filamentary, we conclude that SC is suppressed locally by Cu doping and competes with the CDW in Cu-doped TaSe$_3$. The resistance anomaly due to the CDW transition was extremely small and the size of the anomaly was enhanced with increasing Cu concentration but the temperature at which the anomaly appeared hardly changed. This result of the anomaly and the local suppression of SC imply that the induced CDWs are short-range order in the vicinity of Cu atoms. We also discuss the effect of the pinning of CDWs on the relationship between SC and short-range order CDWs.

\end{abstract}


\maketitle


\section{Introduction}
The relationship between superconductivity (SC) and charge density wave (CDW) has been a major research topic in condensed matter physics. For example, in ZrTe$_3$ in which SC and a CDW exist under atmospheric pressure, the application of pressure below 2 GPa enhances the CDW and eliminates the SC transition~\cite{Yomo2005}. On the other hand, the CDW is suppressed above 2 GPa while the SC transition emerges again at $\sim$5 GPa. Moreover, in NbSe$_3$ with a typical CDW material, Ta doping suppresses two CDWs and induces the SC transition~\cite{Kawabata1985}. These results show that the SC or the CDW is enhanced while the other is suppressed, which is ascribed to the competition between SC and the CDW on a Fermi surface. On the other hand, in a Cu$_x$TiSe$_2$ system where Cu atoms are doped in 1$T$-TiSe$_2$ with a commensurate CDW material, a unique SC and CDW phase diagram is observed and a new relationship between SC and CDW has been discussed~\cite{Morosan2006, Novello2017, Kogar2017}. According to the temperature dependence of the resistance, the CDW is suppressed by Cu doping while SC is induced, and the dependence of the SC transition temperature ($T_\mathrm{C}$) on Cu concentration has a domelike structure~\cite{Morosan2006}. However, precise measurements performed with scanning tunneling microscopy (STM) and X-ray diffraction (XRD) reveal that short-range order CDWs survive up to high Cu concentrations where the SC emerges and the transition temperature of the short-range order CDWs is independent of the $T_\mathrm{C}$~\cite{Novello2017, Kogar2017}. This result suggests that short-range order CDWs are not necessarily in competition with SC and cannot be explained solely in terms of the competition for the density of states at the Fermi level. Moreover, according to the XRD result, the short-range order CDWs change from commensurate to incommensurate at the Cu concentration where SC emerges in Cu$_x$TiSe$_2$~\cite{Kogar2017}. The result suggests that the incommensuration of the CDWs may affect the relationship between SC and the CDWs. However, the reason for this is unclear. As described above, the relationship between SC and CDW has been investigated in many materials. However, previous studies were actually limited to two cases: the first is where SC and a CDW intrinsically exist; the second is where SC is induced in a CDW material. Here, in addition to these two cases, we should also investigate the relationship in a third case where a CDW is induced in a superconducting material, if we are to understand the whole picture of the relationship between SC and CDW.

TaSe$_3$ provides a stage on which to investigate the relationship in the third case. TaSe$_3$ is one of the transition metal tri-chalcogenides, MX$_3$ (M: Nb, Ta; X: S, Se). MX$_3$ has a structure consisting of chains made of transition metals and chalcogens, and the chains are weakly bonded by van der Waals forces. Owing to this structure, MX$_3$ is a quasi-one-dimensional conductor in which an electric current travels well in the direction of the chain axis ($b$-axis). Most forms of MX$_3$ (NbSe$_3$, TaS$_3$, and NbS$_3$) exhibit CDW transitions because the Fermi surfaces are flat and the nesting condition is good~\cite{Ong1977, Sambongi1977TaS3, Wang1989, Canadell1990}. On the other hand, TaSe$_3$ exhibits no CDW transition over the entire temperature range~\cite{Regueiro1986}. It is considered that the irregular behavior arises because TaSe$_3$ is far from one-dimensional compared with the other MX$_3$ compounds as you can see from the electrical conductivity anisotropy and the Fermi surfaces~\cite{Geserich1986, Perucchi2004, Takoshima1980, Ong1978, Canadell1990}. Instead of a CDW transition, TaSe$_3$ exhibits an SC transition at $\sim$2 K~\cite{Sambongi1977TaSe3}. The SC transition curve in the temperature dependence of the resistivity is anisotropic~\cite{Morita1987}. The resistivity parallel to the $b$-axis falls to zero at $T_\mathrm{C}$, while that perpendicular to the $b$-axis exhibits a partial drop at $T_\mathrm{C}$ and a finite value even below $T_\mathrm{C}$. Moreover, the diamagnetism induced by the SC transition is not observed in relation to the temperature dependence of the magnetic susceptibility~\cite{Nagata1991}. From these results, it is considered that the SC of TaSe$_3$ is filamentary and composed of superconducting filaments parallel to the $b$-axis.

Recently, we reported changes in the properties of TaSe$_3$ due to Cu doping~\cite{Nomura2017}. The result of single-crystal XRD analysis implies that Cu atoms intercalated in the van der Waals gap change the lattice parameters, leading to an improvement in the nesting condition. Furthermore, we found an anomalous ``$\gamma$''-shaped dip in the temperature derivative of the resistance ($\mathrm{d}R/\mathrm{d}T$) at $\sim$91 K in Cu-doped TaSe$_3$, which is never observed in pure TaSe$_3$. Based on these results, we concluded that a CDW is induced by Cu doping in TaSe$_3$.

Therefore, in this study, we investigated SC in Cu-doped TaSe$_3$ by measuring the temperature dependence of the resistance to clarify the effect of Cu doping on SC and the relationship between SC and an induced CDW.

\section{Experimental}
We prepared single crystals of pure TaSe$_3$ and Cu-doped TaSe$_3$ synthesized by the vapor phase transport method in the same way as our previous study~\cite{Nomura2017}. The growth temperature was 660$^\circ\mathrm{C}$. The nominal value of Cu to Ta was 10\% in Cu-doped TaSe$_3$. Crystals are ribbon-shaped with typical dimensions of $5\,\mu \mathrm{m} \times 10\,\mu \mathrm{m} \times 5\, \mathrm{mm}$. The ribbon plane is ($\overline{2}$01)~\cite{Yamamoto1978}.

\begin{figure}[!h]
 \begin{center}
  \centerline{\includegraphics[width=8cm]{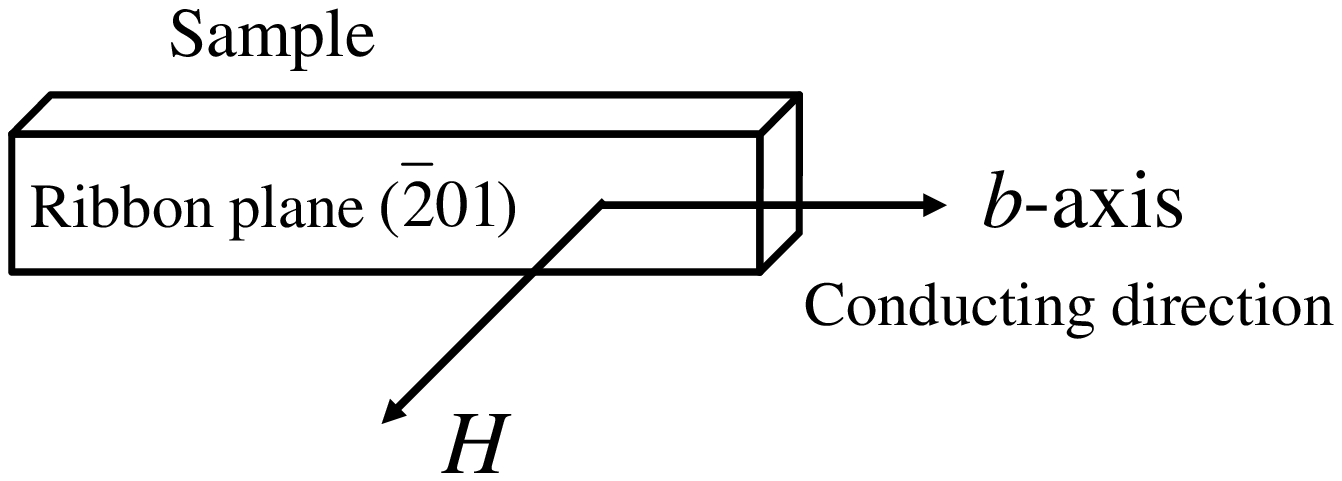}}
 \end{center}
\vspace*{-1cm}
\caption{Schematic of the conduction direction and the magnetic field direction.}
\label{Fig.1}
\end{figure}

The temperature dependence of the resistance along the $b$-axis from 0.6 to 280 K was measured with a dc four-probe technique. The whisker crystal was cut to about 2 mm. Current terminals were attached to both ends of the crystal with carbon paste, and voltage terminals were attached inside. The cross-section at both ends of the crystal was covered with carbon paste. We measured the resistance while the samples were being warmed from 2 to 280 K over about 30 hours and from 0.6 to 2 K over about 30 minutes. The measurements from 0.6 to 2 K were performed applying static magnetic fields (0--80 mT) perpendicular to the ribbon plane after the zero field cooling. Figure \ref{Fig.1} shows a schematic of the conduction direction and the magnetic field direction.

\section{Results}
\begin{figure*}
 \begin{minipage}{0.32\hsize}
  \begin{center}
   \centerline{\includegraphics[width=5.5cm]{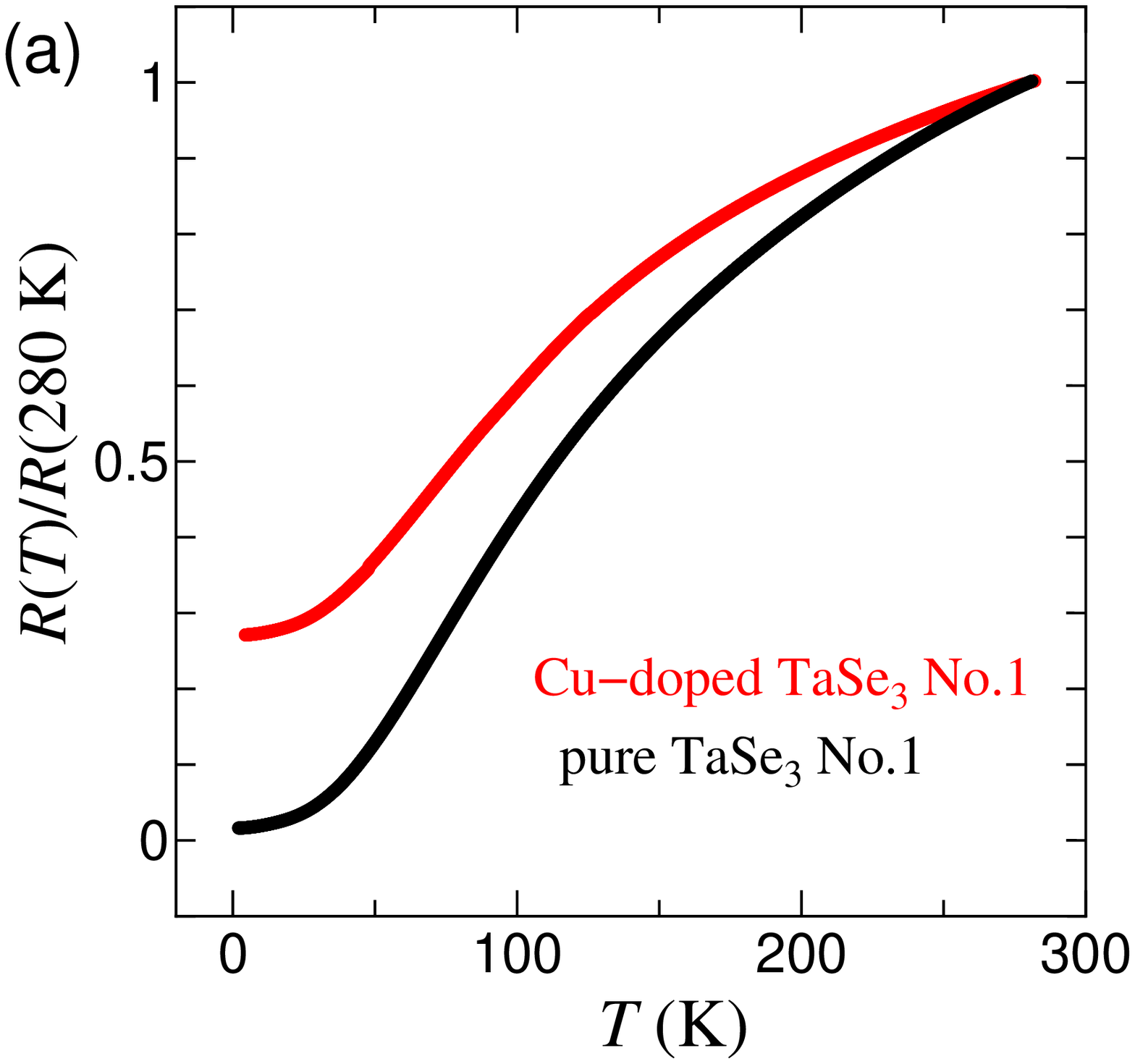}}
  \end{center}
  \end{minipage}
 \begin{minipage}{0.32\hsize}
  \begin{center}
   \centerline{\includegraphics[width=5.5cm]{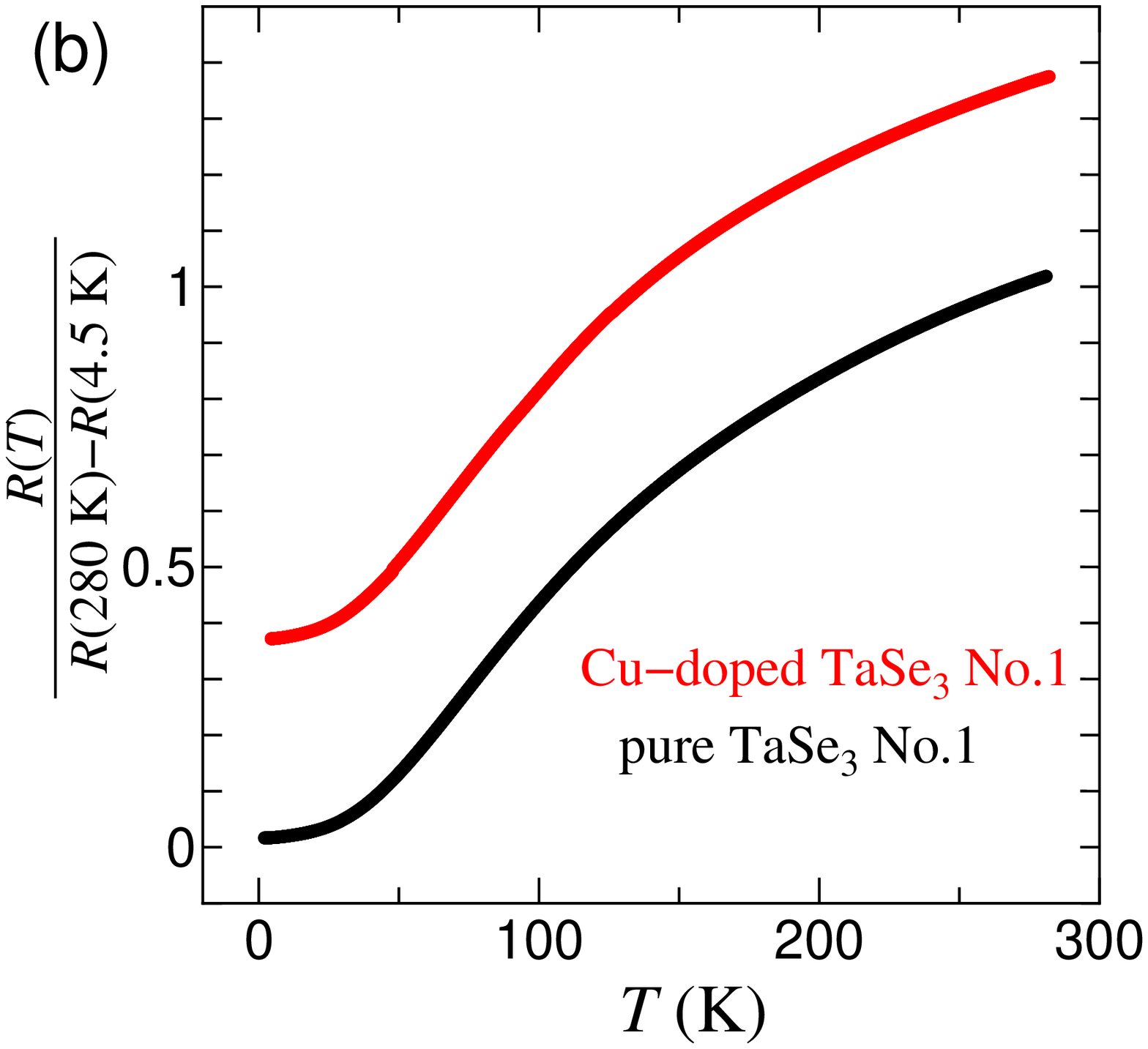}}
  \end{center}
 \end{minipage}
\begin{minipage}{0.32\hsize}
  \begin{center}
   \centerline{\includegraphics[width=5.5cm]{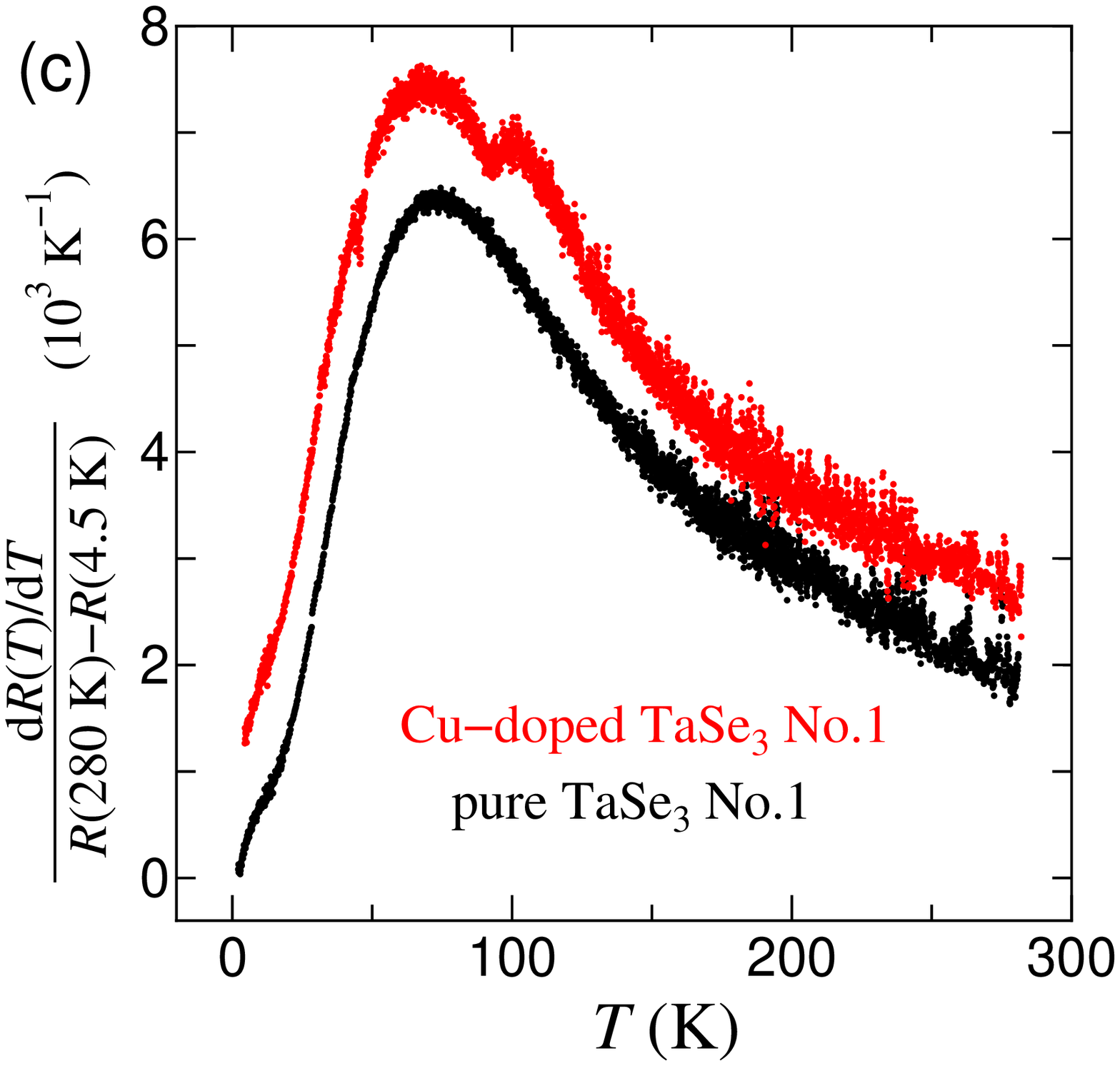}}
  \end{center}
\end{minipage}
\vspace*{-0.5cm}
\caption{Resistance measurement results for of pure TaSe$_3$ and Cu-doped TaSe$_3$ from 2 K to 280 K.
(a) The temperature dependence of the resistance normalized by the resistance at 280 K.
(b) The temperature dependence of the resistance normalized by the difference between the resistances at 280 K and 4.5 K.
(c) The temperature derivative of the normalized resistance in panel (b). The curve of Cu-doped TaSe$_3$ is shifted vertically by 0.001 K$^{-1}$.}
\label{Fig.2}
\end{figure*}

The temperature dependence of the resistance was measured for two samples of pure TaSe$_3$ and for four samples of Cu-doped TaSe$_3$. Figure~\ref{Fig.2} shows a typical result for the resistance measurement from 2 K to 280 K for pure TaSe$_3$ and Cu-doped TaSe$_3$. In accordance with the normalization approaches described in our previous study~\cite{Nomura2017}, the temperature dependence of the resistance was normalized by the resistance at 280 K in Fig.~\ref{Fig.2}(a) and by the difference between the resistances at 280 K and 4.5 K in Fig.~\ref{Fig.2}(b). As with the results in our previous study~\cite{Nomura2017}, pure TaSe$_3$ and Cu-doped TaSe$_3$ showed that resistance had a metallic temperature dependence. The residual resistance ratio ($RRR=R(280~\mathrm{K})/R(4.5~\mathrm{K})$) of Cu-doped TaSe$_3$ was smaller than that of pure TaSe$_3$. The normalized residual resistance ($R(4.5~\mathrm{K})/(R(280~\mathrm{K})-R(4.5~\mathrm{K}))$) of Cu-doped TaSe$_3$ was larger than that of pure TaSe$_3$. The characteristics of all the samples of pure TaSe$_3$ and Cu-doped TaSe$_3$ are summarized in Table~\ref{table1}. As shown in our previous study, Matthiessen's rule holds for the Cu-doped TaSe$_3$ system, and the normalized residual resistance increases almost as a linear function of the Cu concentration determined by inductively coupled plasma atomic emission spectroscopy (ICP-AES)~\cite{Nomura2017}. We obtained the empirical formula
\begin{eqnarray}
y = 0.41 x-0.02,
\label{eq.1}
\end{eqnarray}
where $y$ is the normalized residual resistance and $x$ is the Cu concentration [\%]. Using this empirical formula, we estimated the Cu concentration of each Cu-doped TaSe$_3$ sample from the normalized residual resistance that we observed in this research (Table~\ref{table1}). The estimated Cu concentration varied in the 0.78--1.04\% range among samples from the same batch. Figure~\ref{Fig.2}(c) shows the $\mathrm{d}R/\mathrm{d}T$ in Fig. \ref{Fig.2}(b). A ``$\gamma$''-shaped dip in $\mathrm{d}R/\mathrm{d}T$ was observed at $\sim$91 K in Cu-doped TaSe$_3$. The dip size tended to increase with increasing estimated Cu concentration, which is consistent with the positive correlation between the dip size and the Cu concentration determined by ICP-AES in our previous study~\cite{Nomura2017}.

\begin{table*}
\caption{Characteristics of samples of pure TaSe$_3$ and Cu-doped TaSe$_3$.}
\label{table1}
\begin{center}
\begin{tabular}{lllllll}

\hline
Sample&$R(4.5\,\mathrm{K})~[\Omega]$&$R(280\,\mathrm{K})~[\Omega]$&$RRR$&$y^\ast$&$x$~[\%]$^{\ast\ast}$&$T_\mathrm{C}~[\mathrm{K}]$\\
\hline
pure TaSe$_3$ No.1&2.278&135.391&59.43&0.01711&&1.72\\
pure TaSe$_3$ No.2&1.385&81.256&58.67&0.01734&&1.54\\
Cu-doped TaSe$_3$ No.1&5.613&20.695&3.687&0.3722&0.86&(1.6)$^{\ast\ast\ast}$, 1.07\\
Cu-doped TaSe$_3$ No.2&67.05&217.22&3.240&0.4465&1.04&(1.5)$^{\ast\ast\ast}$, 1.14\\
Cu-doped TaSe$_3$ No.3&9.908&37.116&3.746&0.3642&0.83&1.51, 1.15\\
Cu-doped TaSe$_3$ No.4&11.77&46.58&3.957&0.3382&0.78&1.69\\
\hline
\multicolumn{7}{l}{$\ast$$y$ is the normalized residual resistance defined as $R(4.5\,\mathrm{K})\,/\,(R(280\,\mathrm{K})-R(4.5\,\mathrm{K}))$.}\\
\multicolumn{7}{l}{$\ast\ast$$x$ is Cu concentration.}\\
\multicolumn{7}{l}{$\ast\ast\ast$Approximate temperature when the resistance begins to decrease.}\\

\end{tabular}
\end{center}
\end{table*}

\begin{figure*}
 \begin{minipage}{0.32\hsize}
  \begin{center}
   \centerline{\includegraphics[width=5.5cm]{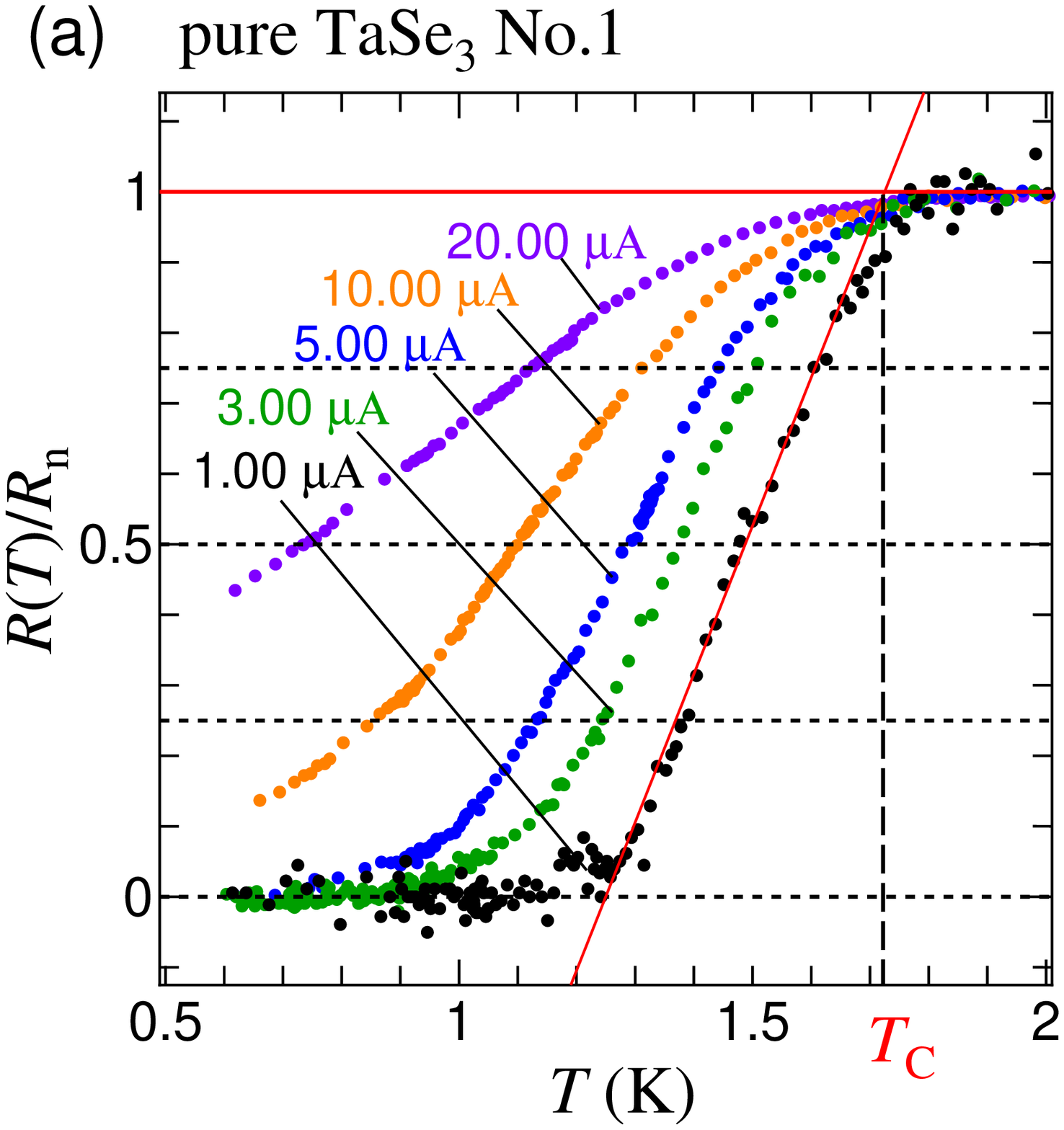}}
  \end{center}
  \end{minipage}
 \begin{minipage}{0.32\hsize}
  \begin{center}
   \centerline{\includegraphics[width=5.5cm]{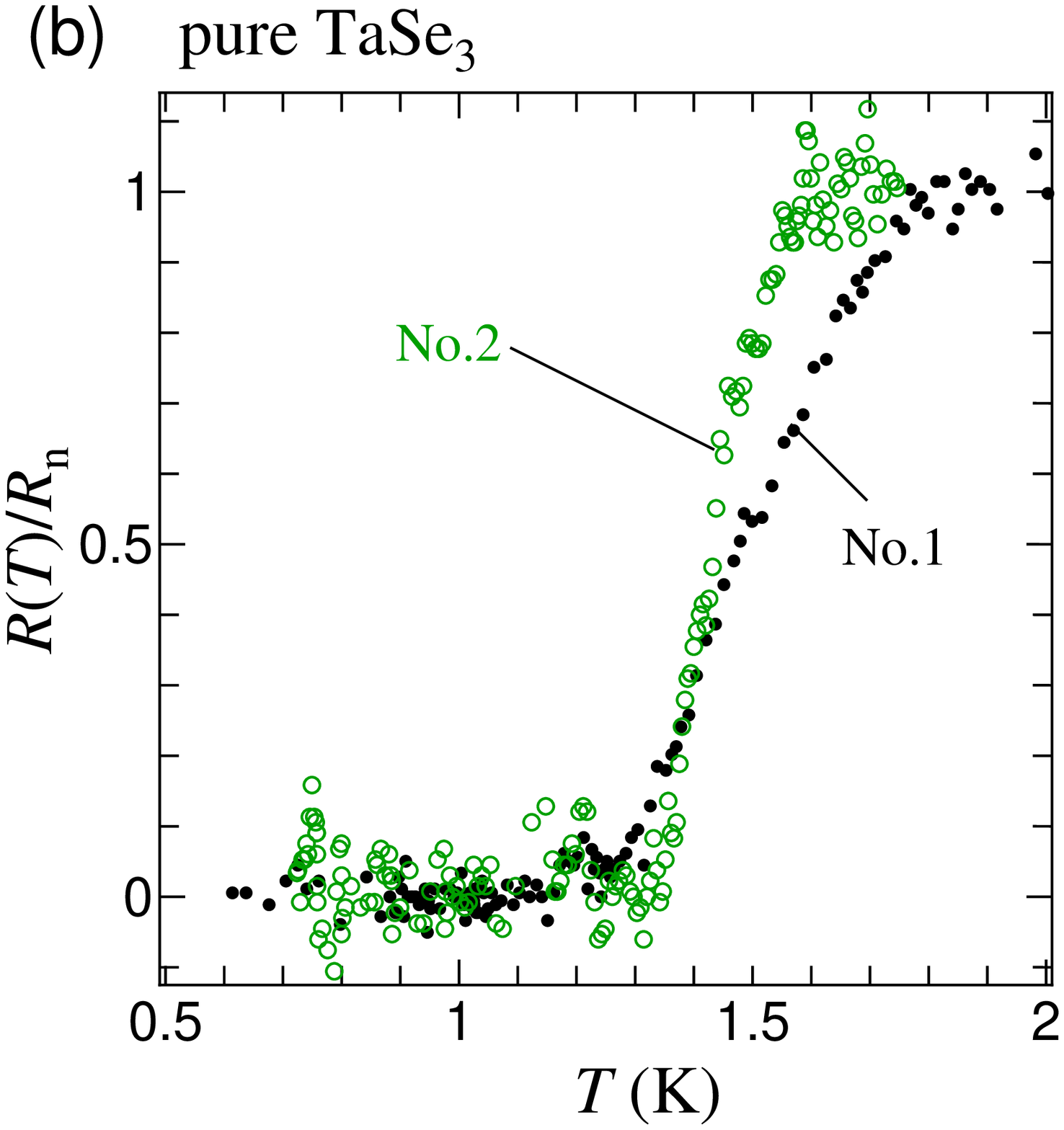}}
  \end{center}
 \end{minipage}
\begin{minipage}{0.32\hsize}
  \begin{center}
   \centerline{\includegraphics[width=5.5cm]{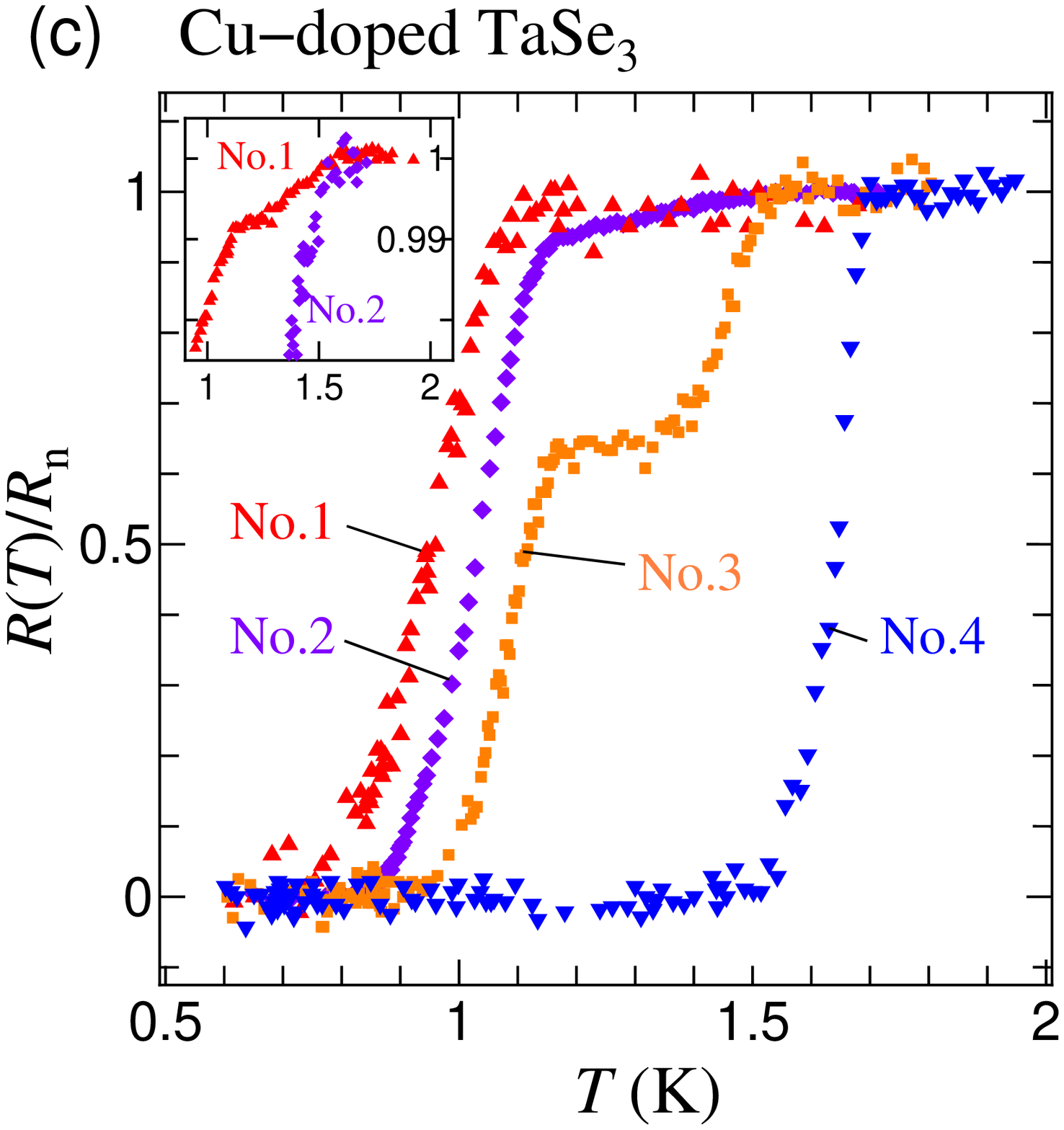}}
  \end{center}
\end{minipage}
\vspace*{-0.5cm}
\caption{The temperature dependence of the resistance from 0.6 K to 2 K normalized by the normal-state resistance ($R_\mathrm{n}$) of pure TaSe$_3$ and Cu-doped TaSe$_3$. 
(a) The SC transition curves for various electric currents of pure TaSe$_3$ sample No.1, and the definition of $T_\mathrm{C}$. We define $T_\mathrm{C}$ as the intersection between a straight line fitted to the data of $R=0.25R_\mathrm{n}$--$0.75R_\mathrm{n}$ and the straight line of $R= R_\mathrm{n}$.
(b) Pure TaSe$_3$ No.1 ($I=1.00\,\mu \mathrm{A}$) and No.2 ($I=0.40\,\mu \mathrm{A}$).
(c) Cu-doped TaSe$_3$ No.1 ($I=0.30\,\mu \mathrm{A}$), No.2 ($I=0.30\,\mu \mathrm{A}$), No.3 ($I=0.30\,\mu \mathrm{A}$), and No.4 ($I=0.30\,\mu \mathrm{A}$). A inset is an enlarged view for Cu-doped TaSe$_3$ No.1 and 2. The data of sample No.1 in the inset is for $I=20.00\,\mu \mathrm{A}$, whose scatter is smaller than that for $I=0.30\,\mu \mathrm{A}$.}
\label{Fig.3}
\end{figure*}

Figure \ref{Fig.3} shows the temperature dependence of the resistance of pure TaSe$_3$ and Cu-doped TaSe$_3$ from 0.6 K to 2 K. All the samples of pure TaSe$_3$ and Cu-doped TaSe$_3$ exhibited a sharp drop in resistance due to the SC transition below 1.7 K. Figure~\ref{Fig.3}(a) shows typical SC transitions for various electric currents. The SC transition exhibited electric current dependence, which matches the results of previous studies~\cite{Haen1978, Nagata1991}. With increasing current, the onset of the SC transition did not move while the offset shifted to the low temperature side. Therefore, we determine $T_\mathrm{C}$ as defined in Fig.~\ref{Fig.3}(a) from the SC transition curve observed for the smallest electric current in this measurement. The results are summarized in Table~\ref{table1}. Pure TaSe$_3$ samples No.1 and 2 exhibited a one-step SC transition with a $T_\mathrm{C}$ of 1.54 K and 1.72 K, respectively. In Cu-doped TaSe$_3$ samples No.1 and 2, the resistance began to decrease at $\sim$1.6 K and 1.5 K, respectively, as shown in the inset of Fig.~\ref{Fig.3}(c), and droped mostly at 1.07 K and 1.14 K. Samples No.3 exhibited a two-step SC transition with $T_\mathrm{C}$s of 1.51 K and 1.15 K. Sample No.4 exhibited a one-step SC transition with a $T_\mathrm{C}$ of $\sim$1.69 K. The ratio of the value of the drop in resistance at $\sim$1.1 K to the normal-state resistance is $\sim$0.99 for No.1, $\sim$0.93 for No.2, $\sim$0.63 for No.3, and 0 for No.4. The samples were No. 4, 3, 1 and 2 in ascending order of Cu concentration as shown in Table~\ref{table1}. Thus, the ratio tended to be higher with increasing Cu concentration.

\begin{figure}
 \begin{minipage}{1\hsize}
  \begin{center}
   \centerline{\includegraphics[width=7cm]{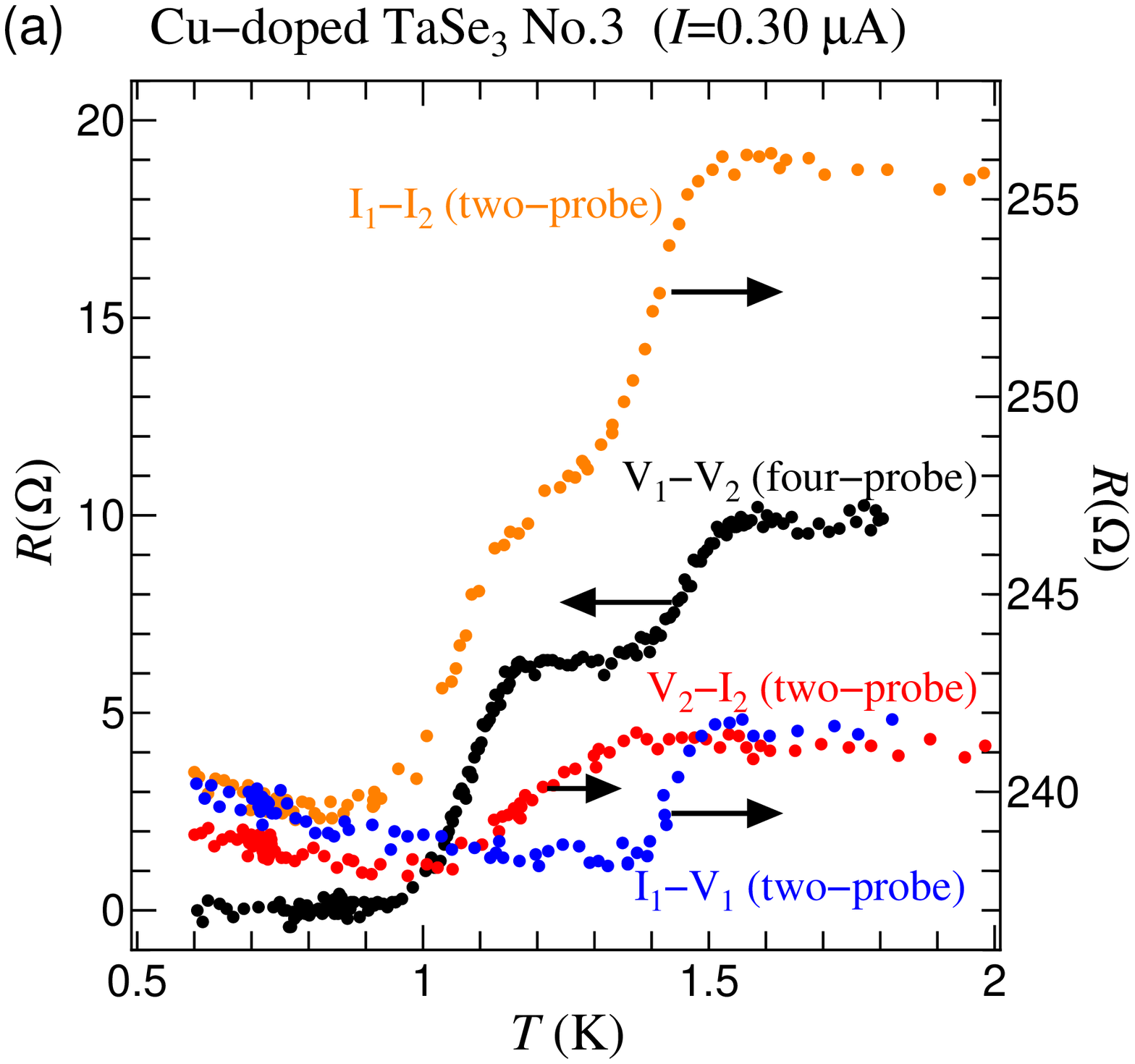}}
  \end{center}
  \end{minipage}
\begin{minipage}{1\hsize}
  \begin{center}
   \centerline{\includegraphics[width=7cm]{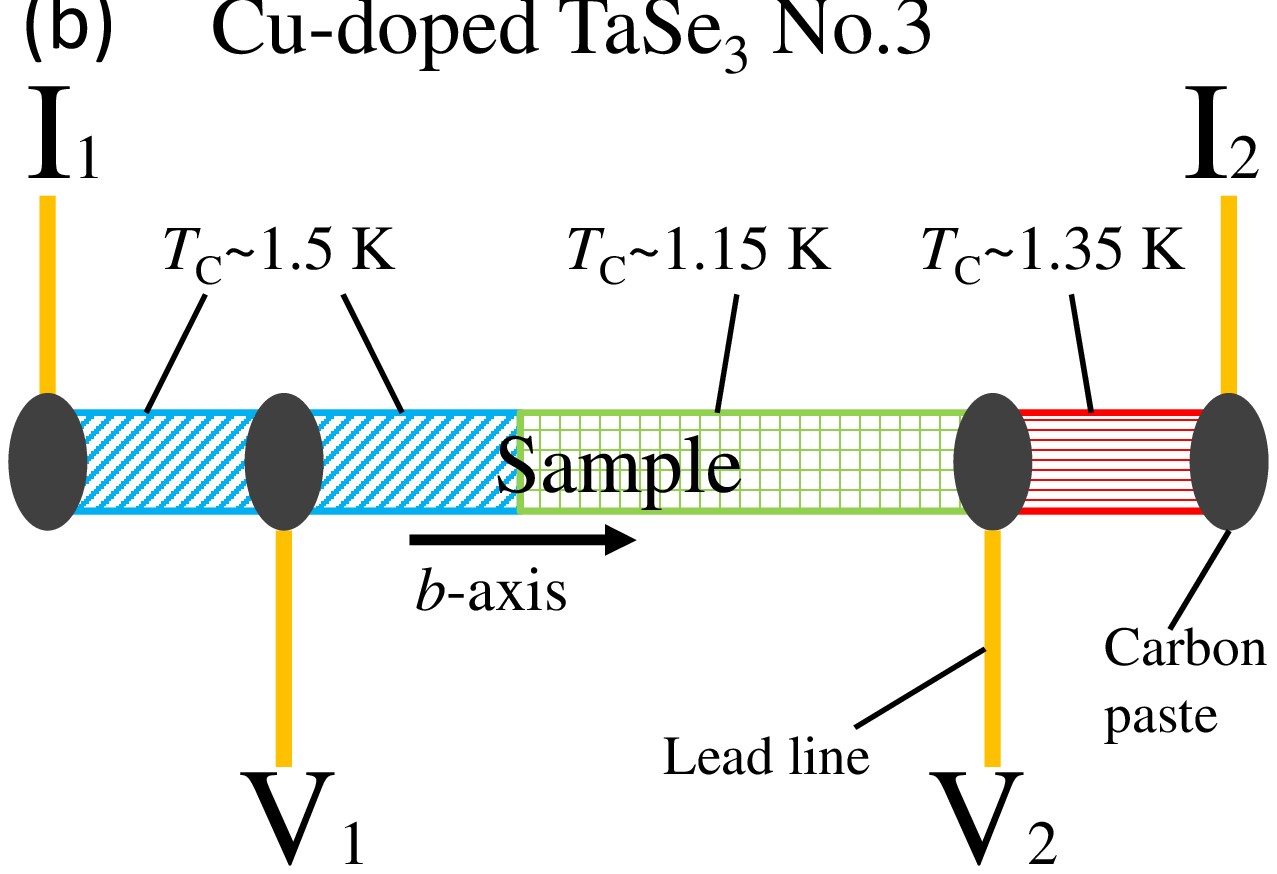}}
  \end{center}
 \end{minipage}
\caption{(a) The temperature dependence of the resistance between the voltage terminals, between the current terminals, and between a current terminal and a voltage terminal in Cu-doped TaSe$_3$ sample No.3. (b) Schematic of spatial distribution of $T_\mathrm{C}$ in Cu-doped TaSe$_3$ sample No.3 derived from panel 4(a). However, the distribution of the regions with the $T_\mathrm{C}$s of $\sim$1.6 K and $\sim$1.15 K between the voltage terminals is deduced.}
\label{Fig.4}
\end{figure}

Cu-doped TaSe$_3$ sample No.3 exhibited a two-step SC transition. To investigate the cause of this result, we measured the temperature dependence of the resistance between the current terminals, and the resistance between a current terminal and a voltage terminal with a dc two-probe measurement. Figure~\ref{Fig.4}(a) shows the results. Sharp drops in resistance due to the SC transition were clearly observed between any pair of terminals, although the resistance measured with a two-probe method did not become zero because of the resistances due to contacts and lead lines ($\sim240~\Omega$). The SC transition differed according to the combination of terminals, showing that various regions with different $T_\mathrm{C}$s are aligned in series in the $b$-axis direction as shown in Fig.~\ref{Fig.4}(b). Thus, the two-step SC transition of sample No.3 was caused by the coexistence of two regions with $T_\mathrm{C}$s of $\sim$1.6 K and $\sim$1.1 K between voltage terminals, and the value of the drop in resistance for each $T_\mathrm{C}$ corresponded to the size of the region with the $T_\mathrm{C}$. Cu-doped TaSe$_3$ samples No.1 and 2 also exhibited a non-single SC transition, where the resistance began to decrease at $\sim$1.6 K and droped mostly at $\sim$1.1 K. Therefore, two regions with $T_\mathrm{C}$s of $\sim$1.6 K and $\sim$1.1 K would coexist and the region with the $T_\mathrm{C}$ of $\sim$1.1 K would occupy most of these samples.

\begin{figure}
 \begin{minipage}{1\hsize}
  \begin{center}
   \centerline{\includegraphics[width=5.5cm]{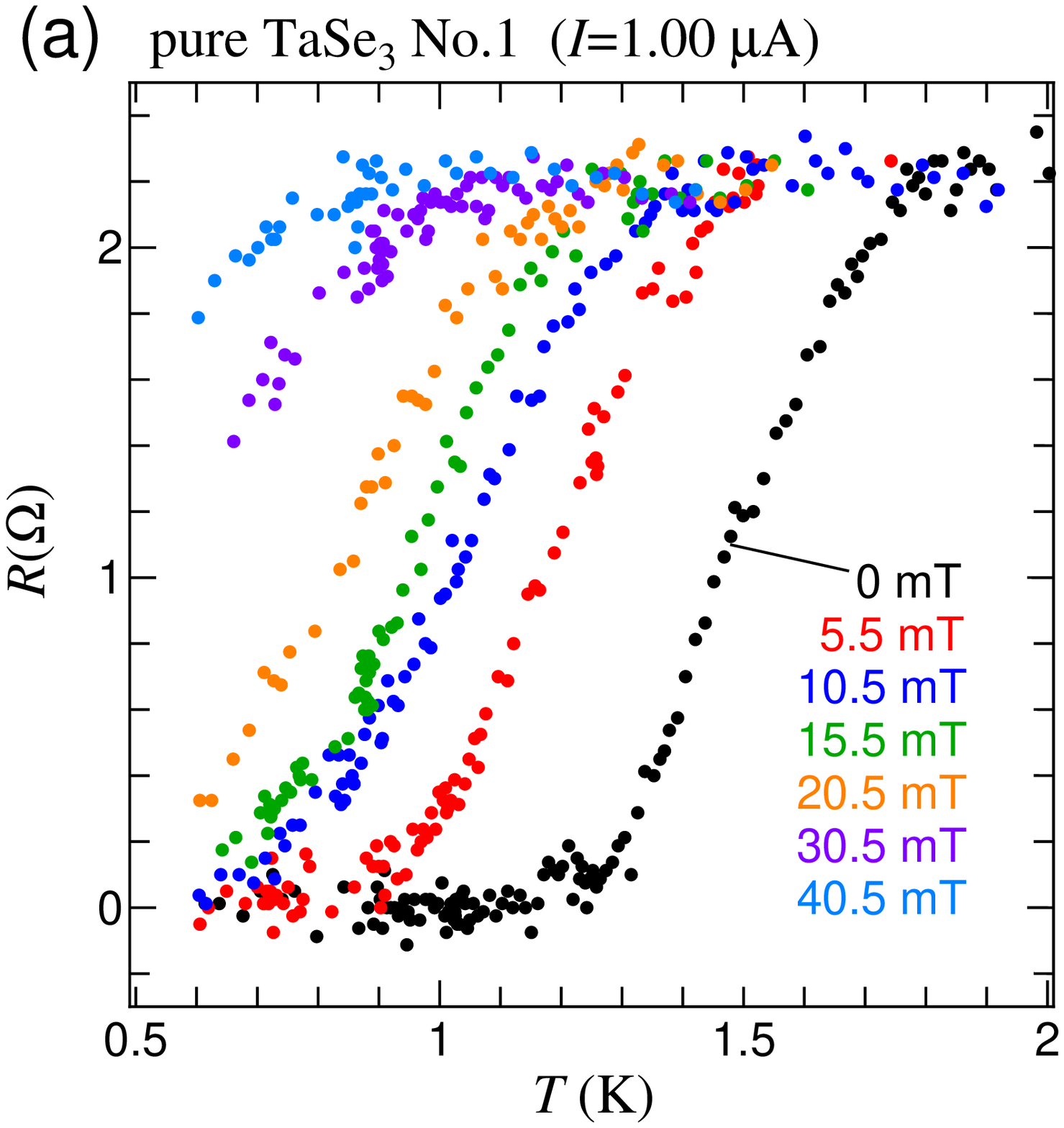}}
  \end{center}
  \end{minipage}
 \begin{minipage}{1\hsize}
  \begin{center}
   \centerline{\includegraphics[width=6cm]{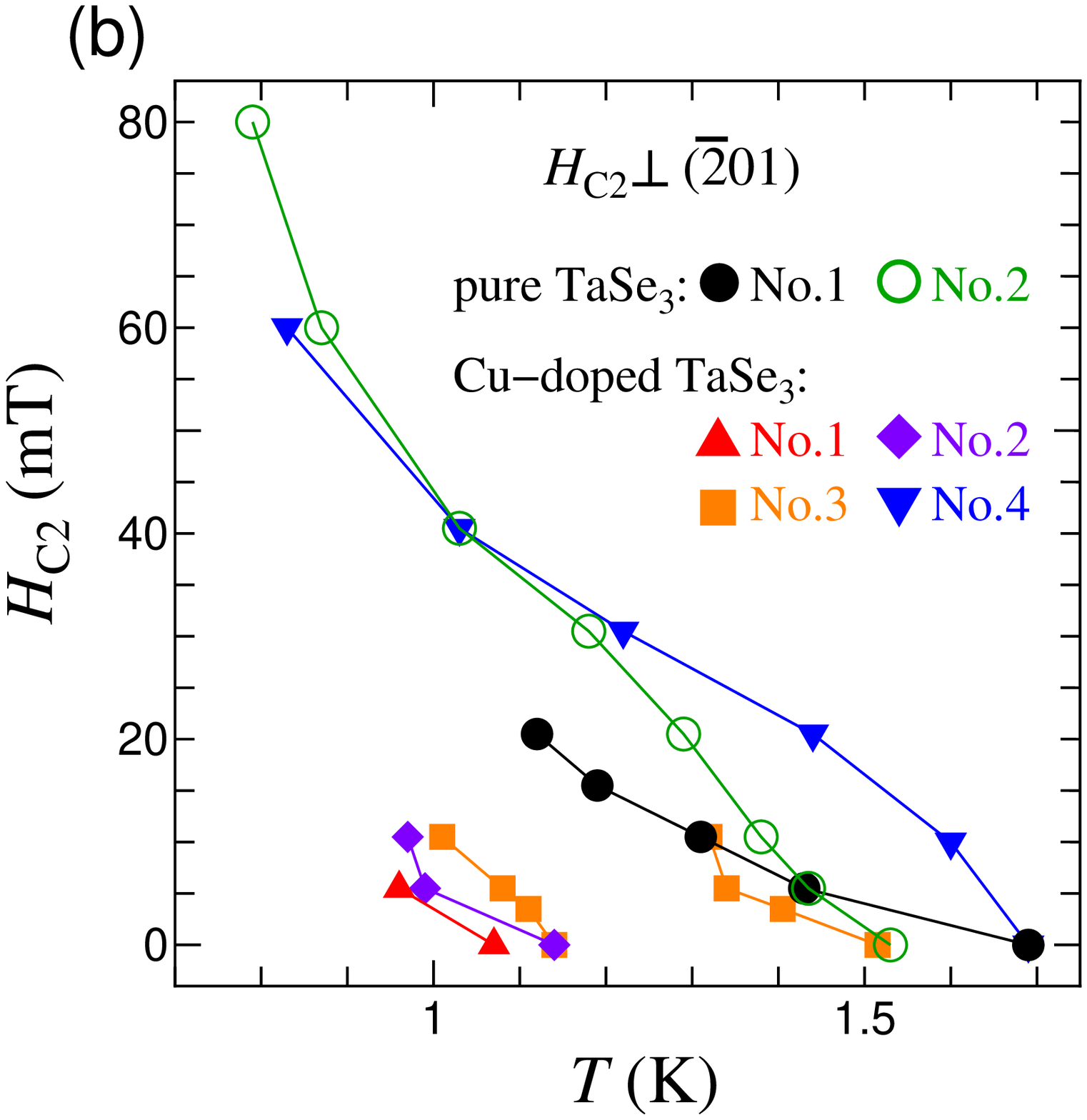}}
  \end{center}
 \end{minipage}
\vspace*{-0.5cm}
\caption{(a) The temperature dependence of the resistance of pure TaSe$_3$ sample No.1 from 0.6 K to 2 K under various magnetic fields perpendicular to the ribbon plane ($\overline{2}$01). 
(b) The temperature dependence of $H_\mathrm{C2}$ perpendicular to the ribbon plane ($\overline{2}$01) of pure TaSe$_3$ and Cu-doped TaSe$_3$.}
\label{Fig.5}
\end{figure}

Figure~\ref{Fig.5}(a) shows a typical temperature dependence of the resistance from 0.6 K to 2 K under various magnetic fields perpendicular to the ribbon plane ($\overline{2}$01). As the magnetic field increased, the SC transition curve shifted to the low temperature side.  Figure~\ref{Fig.5}(b) shows the temperature dependence of the upper critical field ($H_\mathrm{C2}$) perpendicular to the ribbon plane ($\overline{2}$01) of pure TaSe$_3$ and Cu-doped TaSe$_3$. Here we define $H_\mathrm{C2}$ for the $T_\mathrm{C}$ shown in Fig.~\ref{Fig.3}(a). As the temperature decreased, $H_\mathrm{C2}$ of pure TaSe$_3$ sample No.1 increased with an upward curvature, and $H_\mathrm{C2}$ of pure TaSe$_3$ sample No.2 increased with temperature dependence as a linear function near $T_\mathrm{C}$ and with an upward curvature at low temperature. $H_\mathrm{C2}$ of Cu-doped TaSe$_3$ sample No.2 increased with an upward curvature, and $H_\mathrm{C2}$ of Cu-doped TaSe$_3$ sample No.4 increased with a downward curvature near $T_\mathrm{C}$ of $\sim$1.6 K and with an upward curvature at low temperatures. In Cu-doped TaSe$_3$ sample No.3, $H_\mathrm{C2}$ for the first-step SC transition exhibited an upward curvature and that for the second-step SC transition exhibited a slightly downward curvature. The curvature for Cu-doped TaSe$_3$ sample No.1 was unknown because there were only two points.

\section{Discussion}
\begin{figure*}
 \begin{center}
  \centerline{\includegraphics[width=15cm]{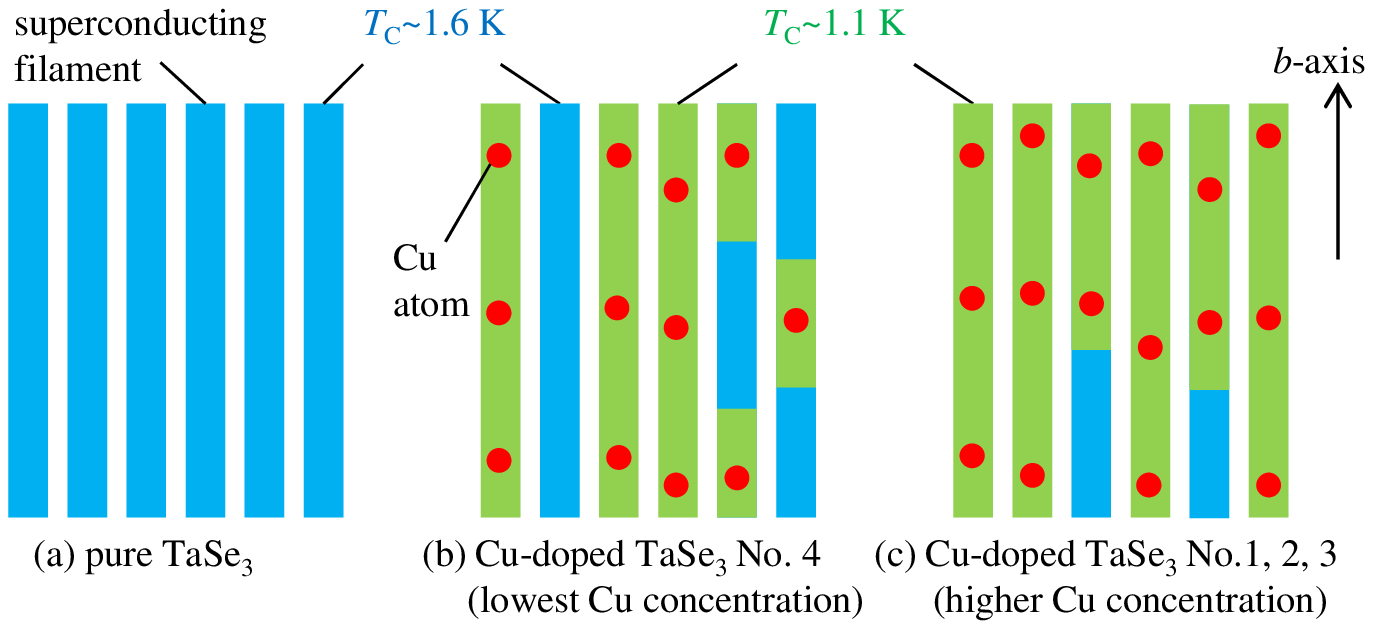}}
 \end{center}
\vspace*{-1cm}
\caption{Model of the superconducting filament structures of pure TaSe$_3$ and Cu-doped TaSe$_3$. The $T_\mathrm{C}$ of with suppressed SC is represented by 1.1 K overall.}
\label{Fig.6}
\end{figure*}

The SC of TaSe$_3$ is filamentary, and adjacent superconducting filaments are coupled to each other by the Josephson effect~\cite{Morita1987, Nagata1991}. An superconducting filament is considered to be composed of multiple TaSe$_3$ chains. In our result, pure TaSe$_3$ exhibits a one-step SC transition at $\sim$1.6 K (Fig.~\ref{Fig.3}(b)). As shown in Fig.~\ref{Fig.6}(a), superconducting filaments with a $T_\mathrm{C}$ of $\sim$1.6 K parallel to the $b$-axis are bundled in pure TaSe$_3$. On the other hand, Cu-doped TaSe$_3$ samples No.1, 2 and 3 with higher Cu concentrations exhibit a coexistence of two regions with $T_\mathrm{C}$s of $\sim$1.6 K, $\sim$1.1 K, and the region with the $T_\mathrm{C}$ of $\sim$1.1 K tends to be larger with increasing Cu concentration (Fig.~\ref{Fig.3}(c) and Fig.~\ref{Fig.4}). This result indicates that the Cu doping suppresses the SC of TaSe$_3$. However, Cu-doped TaSe$_3$ sample No.4 with the smallest Cu concentration exhibits a one-step SC transition at $\sim$1.6 K as with pure TaSe$_3$ (Fig.~\ref{Fig.3}(c)). Moreover, there is a region with a $T_\mathrm{C}$ of $\sim$1.6 K in Cu-doped TaSe$_3$ samples No.1, 2 and 3. The Cu concentrations of the present four Cu-doped TaSe$_3$ samples (See Table~\ref{table1}) are lower than that of the Cu-doped TaSe$_3$ sample with the highest Cu concentration observed in our previous study~\cite{Nomura2017}. The lower Cu concentrations and the location dependence of the $T_\mathrm{C}$ in a sample (Fig.~\ref{Fig.4}) suggest that there are some areas that no Cu atom enters in a sample and the area increases with decreasing Cu concentration. In addition, the suppression of SC by a Cu atom might be limited within a superconducting filament rather than the entire sample because the SC of TaSe$_3$ is filamentary. Therefore, it is believed that, in a sample, there are filaments where Cu atoms enter and SC is suppressed, and filaments where Cu atoms do not enter and SC is not suppressed. If there is a superconducting path composed of superconducting filaments where no suppression remain, the partial suppression of SC cannot be observed by measuring the resistance because the electric current flows through the superconducting path with zero resistance. Therefore, as regards the invariant SC transition in Cu-doped TaSe$_3$ sample No.4 and the region where $T_\mathrm{C}$ does not change in samples No.1, 2 and 3, it can be interpreted that more than one superconducting path with a $T_\mathrm{C}$ of $\sim$1.6 K remains as shown in Fig.~\ref{Fig.6}(b, c).

We compare the temperature dependence of $H_\mathrm{C2}$ for pure TaSe$_3$ and Cu-doped TaSe$_3$. $H_\mathrm{C2}$ is derived from the Pauli paramagnetic and orbital effects. However, the Pauli paramagnetic limit will not matter in the present range of magnetic fields (0--80 mT). Considering orbital effects based on the Ginzburg-Landau (GL) theory, the temperature dependence of the $H_\mathrm{C2}$ of bulk superconductors is expressed as
\begin{eqnarray}
H_\mathrm{C2}(t) = H_\mathrm{C2}(0)\frac{1-t^2}{1+t^2},
\label{eq.2}
\end{eqnarray}
where $t$ is the normalized temperature ($t=T/T_\mathrm{C}$). However, the temperature dependence of the $H_\mathrm{C2}$ of pure TaSe$_3$ and Cu-doped TaSe$_3$ (Fig.~\ref{Fig.5}(b)) cannot be expressed by eq.~\ref{eq.2}. Thus, we require a theoretical model that considers the special feature of SC of TaSe$_3$, namely filamentary SC. According to the GL theory of filamentary SC proposed by L. A. Turkevich and R. A. Klemm (the TK theory), the $H_\mathrm{C2}$ for Josephson-coupled superconducting filaments increases with upward curvature as temperature decreases~\cite{Turkevich1979}. On the other hand, the $H_\mathrm{C2}$ for decoupled superconducting filaments increases with downward curvature. The temperature dependence of the $H_\mathrm{C2}$ exhibits upward curvature in the two pure TaSe$_3$ samples (Fig.~\ref{Fig.5}(b)). From the TK theory, this result implies the Josephson coupling of superconducting filaments. Even in our model as shown in Fig.~\ref{Fig.6}(a), the SC of pure TaSe$_3$ is composed of coupled superconducting filaments with the same $T_\mathrm{C}$. In contrast, the temperature dependence of the $H_\mathrm{C2}$ exhibits downward curvature near a $T_\mathrm{C}$ of $\sim$1.6 K in Cu-doped TaSe$_3$ sample No.4 with the same SC transition as pure TaSe$_3$. Thus, according to the TK theory, the superconducting filaments are decoupled near 1.6 K in this sample. This result is consistent with our model shown in Fig.~\ref{Fig.6}(b) where the superconducting filaments with a $T_\mathrm{C}$ of $\sim$1.6 K are sparse. In other Cu-doped TaSe$_3$ samples No.2 and 3, the temperature dependence of the $H_\mathrm{C2}$ exhibits an upward curvature or a downward curvature. Cu doping results in SC filaments with different $T_\mathrm{C}$s being mixed in a sample as shown in Fig.~\ref{Fig.6}(c). As a result, there are regions where superconducting filaments with the same $T_\mathrm{C}$ are adjacent and coupled, and regions where superconducting filaments with different $T_\mathrm{C}$s are adjacent and not coupled. This complex configuration of superconducting filaments might lead to the complex sample dependence of the $H_\mathrm{C2}$-temperature curve.

From the discussions above, we conclude that the SC of TaSe$_3$ is suppressed by Cu doping. On the other hand, a CDW emerges in Cu-doped TaSe$_3$~\cite{Nomura2017}. Hence, SC and the induced CDW would be in a competitive relationship in Cu-doped TaSe$_3$.

We discuss the induced CDW in Cu-doped TaSe$_3$ because the CDW property is important with discussing the relationship between SC and CDW. The SC suppression of TaSe$_3$ is local in the filament where Cu atoms enter as shown in Fig.~\ref{Fig.6}. Therefore, it is possible that there is no coherent CDW in the entire sample, but CDWs exist locally.

From the viewpoint of the locality of CDWs, we reconsider the resistance anomaly results, that is, the ``$\gamma$''-shaped dip in $\mathrm{d}R/\mathrm{d}T$ due to the CDW transition in Cu-doped TaSe$_3$. Conventional CDW materials such as NbSe$_3$, TaS$_3$ and NbS$_3$ exibit an anomalous increases in resistance at the CDW transition temperatures ($T_\mathrm{CDW}$) because all or part of the Fermi surface disappears and the number of carriers decreases~\cite{Ong1977,Sambongi1977TaS3,Wang1989}. On the other hand, the resistance anomaly of Cu-doped TaSe$_3$ is extremely small~\cite{Nomura2017}. This result for Cu-doped TaSe$_3$ may arise because of the locality of the existence of the CDWs. In fact, in Cu$_x$TiSe$_2$, no signal due to the transition of short-range order CDWs is observed in the temperature dependence of the resistance~\cite{Morosan2006}. Moreover, a comparison of the ``$\gamma$''-shaped dips of Cu-doped TaSe$_3$ samples with different Cu concentrations shows that the dip size increases with increasing Cu concentration while the temperature at which the dip appears is almost constant~\cite{Nomura2017}. The dip size corresponds to the area of the reduced Fermi surface, and the dip appearance temperature corresponds to $T_\mathrm{CDW}$. The relationship between the dip size and the dip appearance temperature appears to contradict the prediction for long-range order CDW with the mean-field theory, which shows a positive correlation between the area of the reduced Fermi surface and $T_\mathrm{CDW}$. However, if the induced CDWs are short-range order in the vicinity of Cu atoms and the correlation length of a CDW is shorter than the distance between Cu atoms, we can interpret the conflicting result mentioned above consistently, namely that the size increases with increasing Cu concentration because the number of short-range order CDWs increases, while the appearance temperature is constant because the state of each short-range order CDW is unchanged.

Here, we estimate the correlation length of short-range order CDWs from a simple calculation of the distance between Cu atoms. The volume of a unit cell of TaSe$_3$ is estimated to be $\sim$0.34 nm$^3$ from the lattice parameters determined in our previous study, and a unit cell contains four Ta atoms~\cite{Nomura2017}. In addition, the Cu concentration of Cu-doped TaSe$_3$ is at most 1\%.  Hence, if Cu atoms enter a crystal evenly, each Cu atom enters a space of $\sim$9 ($=0.34/(4\times0.01)$) nm$^3$ and the distance between adjacent Cu atoms is estimated to be $\sim$2 nm. Therefore, the CDW correlation length must be shorter than $\sim$2 nm.

The estimated size of the CDWs in Cu-doped TaSe$_3$ is nanometer-scale and comparable to the size of the short-range order CDWs in Cu$_x$TiSe$_2$~\cite{Novello2017}. In Cu$_x$TiSe$_2$, the short-range order CDWs do not compete with the SC transition. On the other hand, our results imply that short-range order CDWs suppress SC in Cu-doped TaSe$_3$. One of differences between the two materials is that the CDWs form on Cu atoms in Cu-doped TaSe$_3$ while the CDW is suppressed in the area where Cu atoms enter in Cu$_x$TiSe$_2$~\cite{Novello2017}. Generally, when there is an impurity, a CDW is pinned by that impurity~\cite{Kagoshima1988}. Hence, the CDWs in Cu-doped TaSe$_3$ would be harder to move than those in Cu$_x$TiSe$_2$. Moreover, in Cu$_x$TiSe$_2$, the modulation period of the CDWs changes from commensurate to incommensurate at the Cu concentration where the SC phase emerges~\cite{Kogar2017}. In general, a commensurate CDW is harder to move than an incommensurate CDW~\cite{Kagoshima1988}. Therefore, CDWs may suppress SC when CDWs are hard to move due to impurity pinning or commensurate pinning. We should note that, when discussing the relationship between SC and another electron ordering which is periodically modulated, there is a case where whether the electron ordering is static or dynamic is important~\cite{Tranquada1995}.

\section{Conclusion}
We investigated SC in Cu-doped TaSe$_3$ by measuring the temperature dependence of the resistance. Cu-doped TaSe$_3$ sample with the smallest Cu concentration exhibited a one-step SC transition with a $T_\mathrm{C}$ of $\sim$1.6 K as well as pure TaSe$_3$. On the other hand, those with larger Cu concentrations exhibited a two-step SC transition with $T_\mathrm{C}$s of $\sim$1.6 K and $\sim$1.1 K, or a non-single SC transition where the resistance began to decrease at $\sim$1.6 K and droped mostly at $\sim$1.1 K. The value of the drop in resistance at $\sim$1.1 K tended to expand with increasing Cu concentration. The location dependence of the SC transition showed that the two-step SC transition and the non-single SC transition are caused by the coexistence of two regions with $T_\mathrm{C}$s of $\sim$1.6 K and $\sim$1.1 K aligned in series in the $b$-axis direction in a sample. The temperature dependence of the $H_\mathrm{C2}$ of Cu-doped TaSe$_3$ differed from that of pure TaSe$_3$. From these SC results and the fact that the SC in TaSe$_3$ is filamentary, we conclude that SC is suppressed locally by Cu doping and competes with the CDW in all Cu-doped TaSe$_3$ samples. The resistance anomaly due to the CDW transition was extremely small. The size of the ``$\gamma$''-shaped dip increased with increasing Cu concentration but the temperature at which it appeared hardly changed. These results for the resistance anomaly and the dip, the local suppression of SC imply that the CDWs that appear as a result of Cu doping are in short-range order.

\section{Acknowledgments}
We thank Professor T. Matsuura for helpful discussions. This study is supported by a grant-in-aid for science research from Hokkaido University Clark Memorial Foundation.

\end{document}